\documentclass[%
 reprint,
superscriptaddress,
nofootinbib,
 amsmath,amssymb,
 aps,
prd,
]{revtex4-2}

\usepackage{graphicx}
\usepackage{dcolumn}
\usepackage{bm}

\usepackage[english]{babel}
\usepackage{fancyhdr}
\usepackage{amsmath}
\usepackage{amssymb}
\usepackage{amsfonts}
\usepackage{psfrag}
\usepackage[utf8]{inputenc}
\usepackage[bf,footnotesize,justification=Justified,format=plain]{caption}
\usepackage[dvipsnames]{xcolor}
\usepackage[colorlinks=True, citecolor=blue, linkcolor=blue, urlcolor=blue,linktocpage]{hyperref}
\usepackage{amsthm}
\usepackage{gensymb}
\usepackage{subfig}
\usepackage{physics}
\usepackage{array}
\usepackage{tcolorbox, mathtools}
\usepackage{soul}
\usepackage{aas_macros}
\usepackage[normalem]{ulem}

\newcommand{\mpl}{M_\mathrm{P}}

\newcommand{\centra}{CENTRA, Departamento de Fisica, Instituto Superior T\'ecnico – IST,
Universidade de Lisboa – UL, Avenida Rovisco Pais 1, 1049 Lisboa, Portugal}

\begin{document}

\title{Scalar emission from binary neutron stars in scalar-tensor theories with kinetic screening}


\author{Ramiro Cayuso}
\email{rcayuso@sissa.it}

\affiliation{SISSA, Via Bonomea 265, 34136 Trieste, Italy and INFN Sezione di Trieste}
\affiliation{IFPU - Institute for Fundamental Physics of the Universe, Via Beirut 2, 34014 Trieste, Italy}

\author{Adrien Kuntz}
\email{adrien.kuntz@tecnico.ulisboa.pt}
\affiliation{\centra}

\author{Thiago Assump\c{c}\~{a}o}
\email{tassumpo@uwm.edu}
\affiliation{Center for Gravitation, Cosmology and Astrophysics, Department of Physics \& Astronomy,
University of Wisconsin-Milwaukee, Milwaukee, WI 53211, USA}

\author{Miguel Bezares}
\email{miguel.bezaresfigueroa@nottingham.ac.uk}
\affiliation{Nottingham Centre of Gravity,
Nottingham NG7 2RD, United Kingdom}
\affiliation{School of Mathematical Sciences, University of Nottingham,
University Park, Nottingham NG7 2RD, United Kingdom}

\author{Enrico Barausse}
\email{barausse@sissa.it}

\affiliation{SISSA, Via Bonomea 265, 34136 Trieste, Italy and INFN Sezione di Trieste}
\affiliation{IFPU - Institute for Fundamental Physics of the Universe, Via Beirut 2, 34014 Trieste, Italy}

\date{\today}

\begin{abstract}
We investigate the scalar emission from binary neutron stars in shift-symmetric scalar-tensor theories with kinetic screening ($K$-essence), using 3+1 numerical simulations in the decoupling limit. To construct static binary initial data in the regime where the screening radius $r_*$ greatly exceeds the orbital separation, we introduce a hyperbolization of the static field equations that bypasses the Keldysh-type breakdown affecting direct time evolutions. For equal-mass binaries, where the scalar emission is dominated by the $\ell=m=2$ mode, kinetic screening acts non-monotonically on the scalar radiation, suppressing or enhancing the quadrupolar amplitude depending on the relative size of $r_*$ and $\lambda_{22}$ (with $\lambda_{22}$ the wavelength): for $\lambda_{22}\ll r_*$ it is suppressed relative to the Fierz-Jordan-Brans-Dicke (FJBD) case, while for $\lambda_{22}\gtrsim r_*$ it is amplified above FJBD. For unequal-mass binaries a scalar dipole re-emerges, growing linearly with the mass asymmetry, while the quadrupolar screening remains close to the equal-mass case down to mass ratios $\sim 0.6$. The non-monotonic behavior of kinetic screening that we uncover has potential implications for gravitational-wave-based tests of gravity. The relativistic double pulsar, in particular, requires $r_*\gg 10^9$~km to efficiently suppress the scalar quadrupole; for cosmologically-motivated $\Lambda$, $r_*\sim 10^{11}$~km (for a solar-mass source), giving only moderate suppression.
\end{abstract}

\maketitle

\section{Introduction}

Gravitational-wave (GW) observations of coalescing compact binaries~\cite{TheLIGOScientific:2016src,Abbott:2018lct,LIGOScientific:2019fpa,Abbott:2020jks,LIGOScientific:2021sio} now probe gravity in the strong-field, highly dynamical regime, renewing interest in extensions of General Relativity (GR) that add one or more degrees of freedom beyond the spacetime metric. Among the most studied of such extensions are scalar-tensor theories~\cite{Fierz:1956zz,Jordan:1959eg,Brans:1961sx,Langlois:2015cwa,Crisostomi:2016czh,BenAchour:2016fzp}, in which a single scalar field is coupled non-minimally to gravity and/or matter. Besides serving as theoretical laboratories for testing GR in the strong-field regime, scalar-tensor theories are also motivated on cosmological grounds: suitably chosen models can account for the observed late-time accelerated expansion of the Universe without invoking a cosmological constant~\cite{Clifton:2011jh,Joyce:2014kja,Wolf:2025jed}, while still satisfying stringent Solar-System bounds~\cite{Will:2014kxa}.

A particularly rich such class is that of $K$-essence theories, originally introduced in the inflationary context~\cite{ArmendarizPicon:1999rj} and later exploited for dark-energy scenarios~\cite{Chiba:1999ka,ArmendarizPicon:2000dh}. Their defining feature is a non-canonical kinetic term for the scalar field  $\varphi$ through a generic (nonlinear) function $K(\varphi,X)$ (with $X=g^{\mu\nu}\partial_\mu\varphi\partial_\nu\varphi$), thereby introducing first-derivative scalar self-interactions. A salient feature of some of these theories is that their derivative self-interactions can dynamically switch off modifications of GR in the vicinity of compact or otherwise high-density sources, through the so-called kinetic screening or $k$-mouflage mechanism~\cite{Babichev:2009ee,Kuntz:2019plo,terHaar:2020xxb,Bezares:2021yek,Shibata:2022gec,Lara:2022gof,Boskovic:2023dqk}. In quasi-static, quasi-spherical configurations, kinetic screening confines departures from GR to length scales larger than a characteristic screening radius $r_*$: inside $r_*$ the scalar-matter coupling is effectively suppressed and the dynamics approaches that of GR, whereas outside $r_*$ one recovers the phenomenology of a Fierz-Jordan-Brans-Dicke (FJBD) theory~\cite{Fierz:1956zz,Jordan:1959eg,Brans:1961sx}, i.e.\ a massless scalar-tensor theory in which a single scalar couples universally to matter through a conformal rescaling of the metric, controlled by one dimensionless coupling. This two-regime structure makes these theories particularly appealing from a phenomenological viewpoint, since it potentially reconciles $\mathcal{O}(1)$ scalar-matter couplings at cosmological scales with the tight bounds placed locally by Solar-System and binary-pulsar tests~\cite{Damour:1991rd,Kramer:2006nb,Freire:2012mg}.

The near-luminal propagation of GWs inferred from GW170817~\cite{Monitor:2017mdv,TheLIGOScientific:2017qsa} further restricts the viable landscape of scalar-tensor theories. When this constraint is combined with the requirement that GWs do not decay into dark-energy fluctuations~\cite{Creminelli:2018xsv,Creminelli:2019nok} and with the demand of non-linear stability for the scalar mode~\cite{Creminelli:2019kjy}, the class of viable theories is effectively singled out to be precisely that of $K$-essence with a conformal matter coupling~\cite{Lara:2022gof}. It is therefore of considerable interest to sharpen the observational predictions of this class of theories for compact binary systems.

Realizing this program, however, runs into a well-known difficulty of $K$-essence models: in the fully non-linear regime, the scalar Cauchy problem can become ill-posed~\cite{Babichev:2007dw,Bernard:2019fjb,Bezares:2020wkn,terHaar:2020xxb,Bezares:2021yek,Bezares:2021dma,Lara:2021piy,Shibata:2022gec}. In some situations the scalar equations of motion lose hyperbolicity (``Tricomi-type'' breakdown), whereas in others they formally remain hyperbolic but their characteristic speeds can diverge (``Keldysh-type'' breakdown), preventing any numerically feasible evolution. Several strategies have been devised to handle these obstructions: one can restrict the form of $K(X)$ to subclasses that automatically preserve hyperbolicity~\cite{Babichev:2007dw,Bezares:2020wkn} and thus rule out Tricomi breakdown; one can modify (``fix'') the evolution equations by introducing auxiliary dissipative fields whose impact on secular dynamics is negligible~\cite{Cayuso:2017iqc,Allwright:2018rut,Bezares:2021yek}; one can exploit gauge freedom to bound the characteristic speeds~\cite{Bezares:2020wkn,Bezares:2021dma}; or one can attempt a UV completion of the theory~\cite{Lara:2021piy}, keeping in mind that such a completion may not be compatible with the very existence of the screening mechanism~\cite{Adams:2006sv}. Even so, Keldysh-type pathologies can still arise in physically relevant situations such as gravitational collapse~\cite{Bezares:2020wkn,terHaar:2020xxb,Bezares:2021yek,Akhoury:2011hr,Leonard:2011ce,Gannouji:2020kas}.

These technical advances have made possible the first full numerical-relativity simulations of stellar collapse~\cite{Bezares:2021yek} and of neutron-star binary mergers~\cite{Bezares:2021dma} in $K$-essence with screening. The resulting phenomenology is, however, more subtle than one might have expected. Stellar collapse was found to break kinetic screening, with a substantial amount of scalar radiation being emitted at low frequencies, potentially within the sensitivity band of future space-based interferometers such as LISA~\cite{Bezares:2021yek}. For inspiralling neutron-star binaries, Ref.~\cite{Bezares:2021dma} reported clear suppression of the scalar dipole emission, but an essentially unscreened, or even slightly enhanced, scalar quadrupole. Taken at face value, and extrapolated to the wider range of orbital frequencies probed by binary pulsars, such a result would carry far-reaching implications for the viability of $K$-essence, since an unsuppressed scalar quadrupole would enter radiation reaction at essentially the FJBD level, a regime tightly constrained by binary-pulsar timing. On the other hand, studies of scalar perturbations on static neutron-star backgrounds in $K$-essence~\cite{Shibata:2022gec} and analogous investigations in Galileon scalar-tensor theories~\cite{deRham:2012fg,deRham:2012fw,chu_retarded_2013,andrews_galileon_2013,Dar:2018dra,Brax:2020ujo,deRham:2024xxb} do find efficient suppression of quadrupolar emission, but only once the wavelength of the scalar perturbation is well below $r_*$---a regime that the BNS simulations of Ref.~\cite{Bezares:2021dma} could barely access, given the resolution requirements at realistic cosmologically-motivated values of the strong-coupling scale $\Lambda$, for which $r_*\sim 10^{11}$ km around a solar-mass star.

A first step towards bridging this gap was taken in Ref.~\cite{Cayuso:2024ppe}, which studied neutron-star/black-hole (NS-BH) binaries in $K$-essence in the decoupling limit, i.e.\ evolving only the scalar sector on a flat background while modelling the scalar source with an effective point-particle description---a strategy that had proven effective in Galileon theories~\cite{deRham:2012fg,deRham:2012fw,chu_retarded_2013,andrews_galileon_2013,Dar:2018dra,Brax:2020ujo,deRham:2024xxb}. The decoupling limit allowed Ref.~\cite{Cayuso:2024ppe} to reach separations of scales between the screening radius and the wavelength of the emitted radiation that are inaccessible to full-fledged numerical relativity. In that analysis, the shift symmetry of the underlying $K$-essence theory plays a simplifying role: by known no-hair arguments for stationary black holes in shift-symmetric theories~\cite{Hui:2012qt,Sotiriou:2013qea,Capuano:2023yyh}, the black hole carries no scalar charge, so that only the neutron star sources the scalar field, with the black hole entering only through Kepler's law. Ref.~\cite{Cayuso:2024ppe} analytically demonstrated and numerically confirmed the screening of dipolar radiation in this setup; numerically, the quadrupole was found to be only mildly suppressed (by a factor $\lesssim 3$ at the smallest $\Lambda$ accessible to the simulations) relative to its FJBD value.

Binary neutron-star (BNS) systems, however, occupy a physically distinct region of parameter space, and are among the most observationally relevant sources for constraining $K$-essence theories. On the one hand, they include the binary pulsars that provide some of the most stringent existing tests of radiation reaction in scalar-tensor gravity~\cite{Damour:1991rd,Kramer:2006nb,Freire:2012mg}. On the other hand, in a BNS, the scalar dipolar channel is intrinsically suppressed: unlike the NS--BH case of Ref.~\cite{Cayuso:2024ppe}, in which the black-hole scalar charge vanishes and the two-body charge asymmetry is therefore maximal, a BNS contains two stars of comparable scalar charge, so that the dipole emission is small already at the FJBD level and vanishes identically for identical bodies. The quadrupole thus becomes the dominant scalar-emission channel for near-equal-mass BNSs. For exactly equal masses (and equal sensitivities) the dipole vanishes, isolating the quadrupolar channel; varying the mass ratio switches the dipole back on and lets its impact on the quadrupole be probed. Understanding how kinetic screening acts on these multipoles separately, and as a function of the mass ratio, is thus essential to connect the numerical studies of Refs.~\cite{Bezares:2021dma,Shibata:2022gec,Cayuso:2024ppe} to observations.

In this work, we address this question through 3+1 numerical simulations of the scalar sector in the decoupling limit, sourced by two orbiting compact objects that both carry a non-vanishing scalar charge. We thereby extend the effective-source approach of Ref.~\cite{Cayuso:2024ppe} to the BNS case. A key technical challenge is the construction of static binary initial data in the regime in which the screening radius is much larger than the orbital separation: in that regime, we find that a direct time evolution of the scalar field toward the static configuration is generically plagued by Keldysh-type breakdown developing during the transient, which enforces prohibitively restrictive CFL conditions. To overcome this, we devise a \emph{hyperbolization}~\cite{Ruter:2017iph, Assumpcao:2021fhq} of the static scalar-field equations themselves: the resulting evolution is free of Keldysh pathologies and relaxes smoothly to the desired binary static configuration. Starting from such initial data, we ramp up the orbital angular velocity to its Keplerian value and extract the outgoing scalar radiation.

Our main physical findings can be summarized as follows. For equal-mass binaries, where the scalar emission is dominated by the $\ell=m=2$ mode, the quadrupolar amplitude exhibits a non-monotonic dependence on $\Lambda$: when the wavelength $\lambda_{22}$ of the scalar radiation is smaller than $r_*$, the amplitude is suppressed relative to the FJBD case, with a scaling consistent with $\mathcal{A}\propto r_*^{-4/5}$; when $\lambda_{22}>r_*$, it is instead \emph{amplified} above FJBD, with $\mathcal{A}\propto r_*^{2/5}$.\footnote{We caution that the explicit scaling exponents quoted here have been derived within the cubic $K$-essence truncation adopted in this work [Eq.~\eqref{eq:KX}, with $\beta=0$ and $\gamma=1$], which ensure hyperbolicity of the field equations~\cite{Babichev:2007dw,Bezares:2020wkn}; while we expect the qualitative non-monotonic picture to be robust, the precise exponents may depend on the inclusion of higher-order operators.} These findings reconcile and extend the screening behaviour observed in full numerical-relativity BNS simulations~\cite{Bezares:2021dma} and in perturbative studies of isolated stars~\cite{Shibata:2022gec}; the enhancement regime, however, was not reported in the NS-BH simulations of Ref.~\cite{Cayuso:2024ppe}, where the quadrupole was found to be monotonically suppressed. For unequal-mass binaries, a scalar dipole re-emerges, with an amplitude that grows linearly as the mass ratio departs from unity, while the quadrupolar amplitude decreases quadratically; the screening behaviour of the quadrupole, however, remains close to the equal-mass case for mass ratios larger than $\sim 0.6$. Taken together, our results show that kinetic screening acts non-monotonically on the scalar radiation emitted by compact binaries, suppressing or amplifying the signal depending on the relative size of $r_*$ and $\lambda_{22}$---a feature that should be taken into account when using BNS observations to constrain $K$-essence, whether through binary-pulsar timing or future ground-based gravitational-wave detectors. In the double pulsar PSR~J0737-3039~\cite{Kramer:2006nb,Kramer:2021jcw}, for instance, $\lambda_{22}\sim 10^9$~km, whereas cosmologically-motivated values of $\Lambda$ yield $r_*\sim 10^{11}$~km around a solar-mass source: the system thus lies in the screened region of parameter space, but only moderately so, and the $r_*^{-4/5}$ scaling found above implies a quadrupole emission suppressed relative to FJBD by a few tens.

The rest of this paper is organized as follows. In Section~\ref{sec:theoretical_setup}, we introduce the class of scalar-tensor theories that we consider and describe how binary neutron-star systems are modelled through a two-body effective source, highlighting the differences from the NS-BH case of Ref.~\cite{Cayuso:2024ppe}. Section~\ref{sec:numerical_sol} presents our numerical framework: the hyperbolized solver used to build static binary initial data, the orbital evolution, and the results obtained in the equal- and unequal-mass regimes. We summarize our findings and discuss their observational implications in Section~\ref{sec:conclusions}. Section~\ref{sec:expected_scaling} provides an analytic scaling argument for the equal-mass BNS quadrupole at large distances, complementing the numerical results of Section~\ref{sec:numerical_sol} and accounting for the observed $\mathcal{A}\propto r_*^{-4/5}$ scaling in the regime $\lambda_{22}\ll r_*$. Throughout this paper, we use geometric units in which $c=\hbar=1$.

\section{Theoretical setup} \label{sec:theoretical_setup}

We focus on the shift-symmetric subclass of $K$-essence scalar-tensor theories in which the non-minimal kinetic function depends only on the canonical kinetic term
$X\equiv g^{\mu\nu}\partial_\mu\varphi\,\partial_\nu\varphi$,
while matter fields couple universally to a Jordan-frame metric
$\tilde{g}_{\mu\nu}=A(\varphi)\,g_{\mu\nu}$ whose conformal factor depends only on $\varphi$~\cite{terHaar:2020xxb,Bezares:2021yek,Shibata:2022gec,Kuntz:2019plo,Boskovic:2023dqk}. The action takes the form
\begin{equation}\label{action2}
    S = \int \mathrm{d}^4 x\,\sqrt{-g}\,\bigg[\frac{\mpl^2}{2}\,R + K(X)\bigg] + S_m\!\left[A(\varphi)\,g_{\mu\nu},\Psi_m\right]\,,
\end{equation}
with $\mpl^2=1/(8\pi G)$ [$G$ being Newton's constant] the Planck mass; $R$ the Ricci scalar of the Einstein-frame metric $g_{\mu\nu}$; and $\Psi_m$ collectively denoting the matter fields. We adopt the effective-field-theory (EFT)  expansions
\begin{align} \label{eq:KX}
    &K(X) = - \frac{X}{2} + \frac{\beta}{4 \Lambda^4} X^2 - \frac{\gamma}{8 \Lambda^8} X^3 + {\cal O}(X^4)\,,\\
    &A(\varphi)= \exp\left[2\alpha \frac{\varphi}{\mpl}+{\cal O}(\varphi^2)\right]\,,
\end{align}
truncated at the order displayed. The coefficients $\alpha$, $\beta$ and $\gamma$ are dimensionless numbers of ${\cal O}(1)$, while $\Lambda$ sets the strong-coupling scale of the EFT. Requiring the scalar to mimic dark energy at late cosmological times singles out $\Lambda\sim(H_0\,\mpl)^{1/2}\simeq 2\times 10^{-3}\,\mathrm{eV}$, with $H_0$ the Hubble rate today.

Tricomi-type pathologies of the scalar Cauchy problem are avoided whenever $K(X)$ obeys the hyperbolicity bound~\cite{Babichev:2007dw,Bezares:2020wkn}
\begin{equation}\label{condition}
    1 + 2\,\frac{K''(X)\,X}{K'(X)} > 0\,.
\end{equation}
For the truncation in Eq.~\eqref{eq:KX}, this is satisfied for all $X$ by taking $\beta=0$ and $\gamma>0$; throughout this work we set $\beta=0$ and $\gamma=1$, and we have checked that our conclusions do not depend sensitively on this prescription. The radiative stability of the truncation in the non-linear screened regime has been established in Refs.~\cite{deRham:2014wfa,Brax:2016jjt}.

Varying the action~\eqref{action2} with respect to $\varphi$ gives the scalar equation of motion
\begin{equation} \label{eq:scalarEOM}
    \nabla_\mu \big[ K'(X)\, \nabla^\mu \varphi \big] = \Sigma\,,
\end{equation}
with source
\begin{equation}\label{sigma_generic}
    \Sigma \equiv \frac{1}{2\sqrt{-g}}\,\frac{\delta S_m}{\delta \varphi}\,.
\end{equation}
When matter couples universally and minimally to $\tilde g_{\mu\nu}$ (i.e.\ in the absence of strong-equivalence-principle violations), Eq.~\eqref{sigma_generic} reduces to $\Sigma=(\alpha/2\mpl)\,T$, with
$T = g^{\mu\nu}T_{\mu\nu}$ the Einstein-frame trace of the matter stress-energy tensor
${T}^{\mu\nu}\equiv(2/\sqrt{-g})\,\delta S_m/\delta g_{\mu\nu}$.

State-of-the-art numerical-relativity studies of compact binaries in $K$-essence~\cite{terHaar:2020xxb,Bezares:2021yek,Bezares:2021dma,Shibata:2022gec,Boskovic:2023dqk} evolve the metric jointly with a fluid description of the stars. In the present work we adopt instead the worldline approach used 
routinely both in GR and in modified gravity theories~\cite{Will:1989sk,Damour:1992we,deRham:2012fg,deRham:2012fw,Dar:2018dra,deRham:2024xxb,Brax:2020ujo,Yagi:2013ava,Will:2018ont} and employed for the neutron star--black hole problem in Ref.~\cite{Cayuso:2024ppe}: each neutron star is replaced by an effective point particle, so that the matter sector is described by the two-body action
\begin{equation}\label{ppaction}
    S_m = -\sum_{i=1}^{2}\int \tilde m_i(\varphi)\,d\tilde{\tau}_i\,,
\end{equation}
where $i=1,2$ labels the two stars and the proper time along each worldline is measured with respect to the Jordan-frame metric, i.e.
\begin{equation}
    d\tilde\tau_i = e^{\alpha\varphi/\mpl}\,\sqrt{-g_{\mu\nu}\,dx^\mu\,dx^\nu}\,.
\end{equation}
Inside a compact object, the scalar field renormalizes the effective conformal coupling by an amount that reflects the body's strong-field structure; this is encoded in the explicit $\varphi$-dependence of $\tilde m_i$. The resulting violations of the strong equivalence principle are parametrized by the dimensionless sensitivities~\cite{1975ApJ...196L..59E,Will:1989sk,Damour:1992we,Kuntz:2024jxo}
\begin{equation}
    s_i \equiv -\left.\frac{\partial \ln \tilde m_i}{\partial \ln A}\right|_{\varphi=\varphi_0}\,,
\end{equation}
evaluated on the background scalar configuration $\varphi_0$.

Inserting Eq.~\eqref{ppaction} into Eq.~\eqref{sigma_generic} and carrying out the functional derivative, the scalar source becomes a sum over the two bodies of Dirac delta distributions weighted by the individual scalar charges,
\begin{equation}\label{sigmaused}
    \Sigma = -\frac{1}{2\mpl}\sum_{i=1}^{2}\frac{\alpha_i\,m_i}{\sqrt{-g}\,u_i^{\,t}}\,\delta^{(3)}\!\left(\boldsymbol{x}-\boldsymbol{x}_i(t)\right)\,,
\end{equation}
where $m_i\equiv \tilde m_i\,e^{\alpha\varphi/\mpl}$ are the Einstein-frame masses, $u_i^\mu$ the Einstein-frame four-velocities, $\boldsymbol{x}_i(t)$ the body positions, and
\begin{equation}\label{eq:scalarcharge}
    \alpha_i \equiv \alpha\,(1-2\,s_i)
\end{equation}
is the Einstein-frame scalar charge of the $i$-th star~\cite{Damour:1992we}. The sensitivity-independent expression $\Sigma=(\alpha/2\mpl)\,T$ obtained above is recovered in the $s_i\to 0$ limit of Eq.~\eqref{sigmaused}.

  In Ref.~\cite{Cayuso:2024ppe}, which considered black-hole--neutron-star binaries, the sum in Eq.~\eqref{sigmaused} reduced to a single term, since the shift symmetry of $K$-essence forces
   the black-hole scalar charge to vanish~\cite{Hui:2012qt,Sotiriou:2013qea,Capuano:2023yyh} (equivalently, $s_\mathrm{BH}=1/2$ and $\alpha_\mathrm{BH}=0$), and the black-hole mass therefore
   entered the scalar dynamics only through Kepler's law. Here, by contrast, both bodies source the scalar field, each with its own mass $m_i$ and sensitivity $s_i$, and with scalar charges
  $\alpha_1\neq\alpha_2$ unless the two stars are identical. Note, however, that the sensitivities and the (Einstein-frame) masses enter the scalar source in Eq.~\eqref{sigmaused} only      
  through the products $\alpha_i m_i = \alpha\,(1-2 s_i)\,m_i$. At fixed orbital kinematics---i.e.\ fixed angular frequency $\Omega$ and body positions $\boldsymbol{x}_i(t)$---the scalar
  problem is therefore invariant under the rescaling $m_i\to (1-2 s_i)\,m_i$, which reabsorbs the sensitivities into effective scalar masses. For this reason, and because efficient local
  kinetic screening around compact stars suggests that neutron-star sensitivities in $K$-essence are small to begin with~\cite{terHaar:2020xxb,Bezares:2021yek}, in the numerical simulations
  presented below we set $s_1=s_2=0$ (so that $\alpha_1=\alpha_2=\alpha$) without loss of generality: our results apply to arbitrary pairs of neutron-star sensitivities through the rescaling
   above.

Finally, as in Ref.~\cite{Cayuso:2024ppe}, we work in the decoupling limit: the background metric $g_{\mu\nu}$ entering Eqs.~\eqref{eq:scalarEOM} and~\eqref{sigmaused} is set to Minkowski, and gravitational dynamics is retained only through the Keplerian relation between the orbital angular frequency $\Omega$ and the binary separation $a=a_1+a_2$,
\begin{equation}\label{eq:defOmega}
    \Omega = \sqrt{\frac{G\,M}{a^3}}\,,
\end{equation}
where $a_i$ denotes the distance of the $i$-th star from the binary's center of mass, and $M=m_1 +m_2$ is the total mass of the system.

\subsection{Expected scaling of the quadrupole}\label{sec:expected_scaling}

In this section, we present an analytic argument that predicts the $\mathcal{A}\propto r_*^{-4/5}$ scaling of the equal-mass BNS quadrupolar amplitude in the regime $\lambda_{22}\ll r_*$, which will be confirmed by the numerical results of Section~\ref{sec:numerical_sol}.

Analytically, we have not been able to fully solve for the scalar field dynamics in the BNS case. This is because the approach used in Ref.~\cite{Cayuso:2024ppe}, which allowed us to obtain the scalar field in the near zone of the binary, is not easily generalized to the BNS case. Without a near-zone solution to the scalar equation of motion, we cannot perform the matching computation developed in Ref.~\cite{Cayuso:2024ppe}, which is essential to analytically obtain the amplitude of the scalar at infinity.

This difficulty, however, does not prevent us from obtaining relevant solutions to the scalar field at distances far from the source, which is the purpose of this section. We will not predict the exact amplitude of the scalar, but its scaling with the parameters of the problem can be obtained by analogy with our previous computation. At large distances from the source ($|\boldsymbol{x}|\gg a$), one can still use the same ansatz as in Ref.~\cite{Cayuso:2024ppe} to perturbatively solve for the scalar as a static background component plus a small perturbation:
\begin{equation}
\varphi(t,\boldsymbol{x}) = \varphi_\mathrm{SS}(|\boldsymbol{x}|) + \varphi_1(t, \boldsymbol{x}) \; ,
\end{equation}
where $\varphi_\mathrm{SS}$ is the spherically symmetric field generated by the total mass $M$ of the binary, solving
\begin{equation}
    \frac{1}{r^2} \partial_r \big( r^2 K'((\varphi_\mathrm{SS}')^2) \varphi_\mathrm{SS}' \big) = - \frac{\alpha M}{8 \pi r^2 \mpl} \delta \big( r \big) \; .
\end{equation}
Let us consider an equal-mass BNS. By symmetry, there is no dipolar emission so that $\varphi_1$ is quadrupolar at lowest order. We expand it in Fourier modes and spherical harmonics as
\begin{equation}
\varphi_1 = e^{-2 i \Omega t} \sum_m \mathcal{A}_m R(r) Y_{2m}(\theta,  \phi) + \mathrm{c.c.} \; ,
\end{equation}
where c.c. denotes complex conjugation. In the following, we will focus on the $m=2$ perturbations.
Substituting this ansatz in the scalar equation of motion~\eqref{eq:scalarEOM} and linearizing in $\varphi_1$, we obtain the same equation as in Ref.~\cite{Cayuso:2024ppe},
\begin{equation} \label{eq:EOMR}
    \frac{1}{r^2} \partial_r \big[ r^2(K' + 2 X_\mathrm{SS}K'') R'  \big] + K' \bigg( \omega^2 - \frac{6}{r^2}\bigg) R = 0\,,
\end{equation}
this time for the quadrupolar solution $\ell=2$, with $\omega=2\Omega$ the frequency of the $\ell=m=2$ mode.
At distances much larger than both the screening radius $r_*$ and the quadrupolar wavelength $\lambda_{22}=\pi/\Omega$, the solution to Eq.~\eqref{eq:EOMR} is
\begin{equation} \label{eq:homogeneousSolLargeDistance}
R(r) \simeq \frac{e^{2 i \Omega r} }{r \Omega^{1/2}} e^{-1.12 \Omega r_*} \; ,
\end{equation}
where the constant $1.12$ comes from a numerical integral (see Ref.~\cite{Cayuso:2024ppe} for details). We have therefore solved for the scalar quadrupole at large distances from the source up to a free amplitude $\mathcal{A}_2$. We expect the matching to proceed similarly to that of Ref.~\cite{Cayuso:2024ppe}, with the notable difference that a quadrupolar field is further suppressed by a factor $a\Omega$. This implies that, up to an order-one constant $C$,
\begin{equation} \label{eq:amplitudePrediction}
\mathcal{A}_2 = C\, r_*^{-4/5}\, \frac{M}{\mpl}\, a^2\, \Omega^{17/10} \; ,
\end{equation}
obtained under the assumption that the screening radius is much larger than the quadrupolar wavelength, $r_*\gg\lambda_{22}$.
This explains the scaling of the scalar amplitude at infinity that we observe in the numerical simulations of Section~\ref{sec:numerical_sol}---in particular, the proportionality $\mathcal{A}\propto r_*^{-4/5}$ in the deep-screening regime $\lambda_{22}<r_*$.
Finally, note that this argument relies on the lowest-order scalar perturbation being quadrupolar, which is the case for an equal-mass BNS. For unequal masses, there could be an interplay between dipolar perturbations sourcing a quadrupolar field, which we leave to future work.

\section{Numerical Simulations} \label{sec:numerical_sol}

In this section, we present our numerical simulations. We first construct the static binary configurations and then use them as initial data for the subsequent orbital (Keplerian) evolution.
We start from equal-mass binaries---in which symmetry allows the quadrupolar screening to be studied in isolation from dipolar emission---and then turn to systems with varying mass ratio.


 \subsection{Numerical Scheme}
The numerical scheme is the same as that used in Ref.~\cite{Cayuso:2024ppe}.
The evolution code is a Simflowny-based 3D code  \cite{ARBONA20132321,ARBONA2018170,PALENZUELA2021107675,simflowny2021} which is parallelized and possesses adaptive mesh refinement (AMR) through the SAMRAI infrastructure \cite{https://doi.org/10.1002/cpe.652,GUNNEY201665,SAMRAIurl}. Fourth-order finite difference operators are used to discretize the equations and a fourth-order Runge-Kutta time integrator is implemented for the evolution. The computational domain lies in the range $[-12500,12500]^{3}\,\mathrm{km}^{3}$, containing 10 refinement levels. Each level has twice the resolution of the previous one, achieving a resolution of $\Delta x_{10}= 0.123\,\mathrm{km}$ on the finest grid. We use a Courant factor $\lambda_{c} \equiv \Delta t_{l}/\Delta x_{l} = 0.3$ on each refinement level $l$.

 \subsection{Sources}
The theoretical framework introduced in Sec.~\ref{sec:theoretical_setup} describes the matter sector through an effective point-particle action. While this description is sufficient to analytically derive expected scaling of the quadrupole, a direct numerical implementation of point-like sources would require resolving the singular behavior of the source terms. For this reason, in the numerical simulations we replace each point particle by a smooth matter distribution that approximates the neutron star. This regularization is purely numerical and does not modify the long-wavelength scalar radiation, which is the quantity of interest in this work. The validity of this approach was investigated in detail in Ref.~\cite{Cayuso:2024ppe}, where we demonstrated that different choices of smooth source profiles lead to indistinguishable radiative observables while substantially reducing the numerical resolution requirements.

 To model the matter sources corresponding to the neutron stars, we adopt the same approach as in Ref.~\cite{Cayuso:2024ppe}, where the energy density is given by the superposition of two smooth spherical profiles as follows,
 \begin{equation}\label{eq:matprofile} 
    \rho(t,x,y,z)= \sum_{i=1}^{2} \frac{A_{i}\,\tilde{r}_{i}^{2}}{(2\pi \sigma)^{3/2}}\exp(-\left(\frac{\tilde{r}_{i} - r_{p}}{\sigma}\right)^{2}),
\end{equation}
where $\tilde{r}_{i} =  \sqrt{x_{i}(t)^2 + y_{i}(t)^2 + z(t)^2}$, with $x_{i}(t) = x + (-1)^i a_{i} \cos{(\Omega t)}$, $y_{i}(t) = y + (-1)^i a_{i} \sin{(\Omega t)}$ and $z_{i}(t) = z$,  where $a_{i}$ is the distance of each star to the center of mass of the binary system, constrained by $m_1 a_1 = m_2 a_2$. The normalization constants $A_{i}$ fix the individual masses of each star. The value of $\sigma$ is fixed such that $99 \%$ of the mass of each star is contained within a stellar radius $r_s$, which for simplicity will be equal for both stars. The parameter $r_p$ is used to adjust the shape of the shell-like profile.
As shown in Ref.~\cite{Cayuso:2024ppe}, this shell-like profile relaxes the resolution required to evolve stellar solutions, while leaving the outgoing radiation unaffected.

 \subsection{Initial conditions: Binary static configuration}\label{sec:ID}

Finding the static scalar field configuration for a binary system is considerably more challenging than in the isolated star case. Numerical solutions and analytical approximations were presented in \cite{Kuntz:2019plo,Boskovic:2023dqk}; however, in those cases the binary separation in the Keplerian orbit was not sufficiently small compared to the screening radius to suppress scalar radiation, and therefore not within one of the regimes of interest considered in this work (i.e.\ $\lambda_{22} \ll r_*$).

In principle, since we ultimately evolve the system of equations in time, one could begin from an arbitrary field configuration and allow the system to relax toward the solution dictated by the static matter source. In practice, however, we find that this approach can lead to pathological behavior. Due to the characteristic structure of the equations, the field may develop very large propagation speeds, triggering a Keldysh-type breakdown and imposing extremely restrictive Courant-Friedrichs-Lewy (CFL) conditions to maintain numerical stability. This issue persists even with more informed initial conditions, such as a superposition of single-star static solutions, and becomes increasingly severe as $\Lambda$ and the binary separation $a$ decrease.

To overcome these difficulties, we instead adopt a hyperbolization of the static field equations, whereby an elliptic partial differential equation (PDE) is transformed into a hyperbolic PDE for relaxation~\cite{Ruter:2017iph,Assumpcao:2021fhq}. Unlike the original evolution system, the resulting set of equations has a well-behaved characteristic structure and is free from Keldysh-type instabilities. This allows one to obtain the desired static configuration by evolving from arbitrary initial data in a stable manner. The static limit of Eq. \eqref{eq:scalarEOM} is given by:

\begin{equation}
     0 =  \partial_{i}\left(K^{\prime}(\widetilde{X}) \psi^{i} \right) -\Sigma,
\end{equation}
where we define $\psi_{i} \equiv \partial_{i}\varphi $ and $\widetilde{X} = \psi_{i}\psi^{i}$. The hyperbolized version of this last equation reads:

\begin{equation}
    \label{eq:hyperbolized_ID}
     \partial_{\tau}^2\varphi + \eta \partial_{\tau}\varphi =  \partial_{i}\left(K^{\prime}(\widetilde{X}) \psi^{i} \right) -\Sigma,
\end{equation}
where we have introduced a relaxation time $\tau$ and a damping parameter $\eta$. Through a combination of wave-like error propagation and numerical dissipation, the relaxation drives the hyperbolized equation toward the solution of the original elliptic system as both $\partial_{\tau}^2\varphi$ and $\partial_{\tau}\varphi$ approach zero in the steady-state regime. In practice, we perform the relaxation by numerically solving Eq.~\eqref{eq:hyperbolized_ID} as a first-order system in time and space:
\begin{align}
    &\partial_{\tau}\widetilde{\Psi} =  \partial_{i}\left(K^{\prime}(\widetilde{X}) \psi^{i} \right) -\Sigma~,\\
    &\partial_{\tau}\psi_{i} = \partial_{i}\widetilde{\Psi} -\eta \psi_{i}~,\\
    &\partial_{\tau} \varphi = \widetilde{\Psi} -\eta \varphi~.
\end{align}

\begin{figure*}[t]
    \centering



    \centering
    \includegraphics[width=1\linewidth]{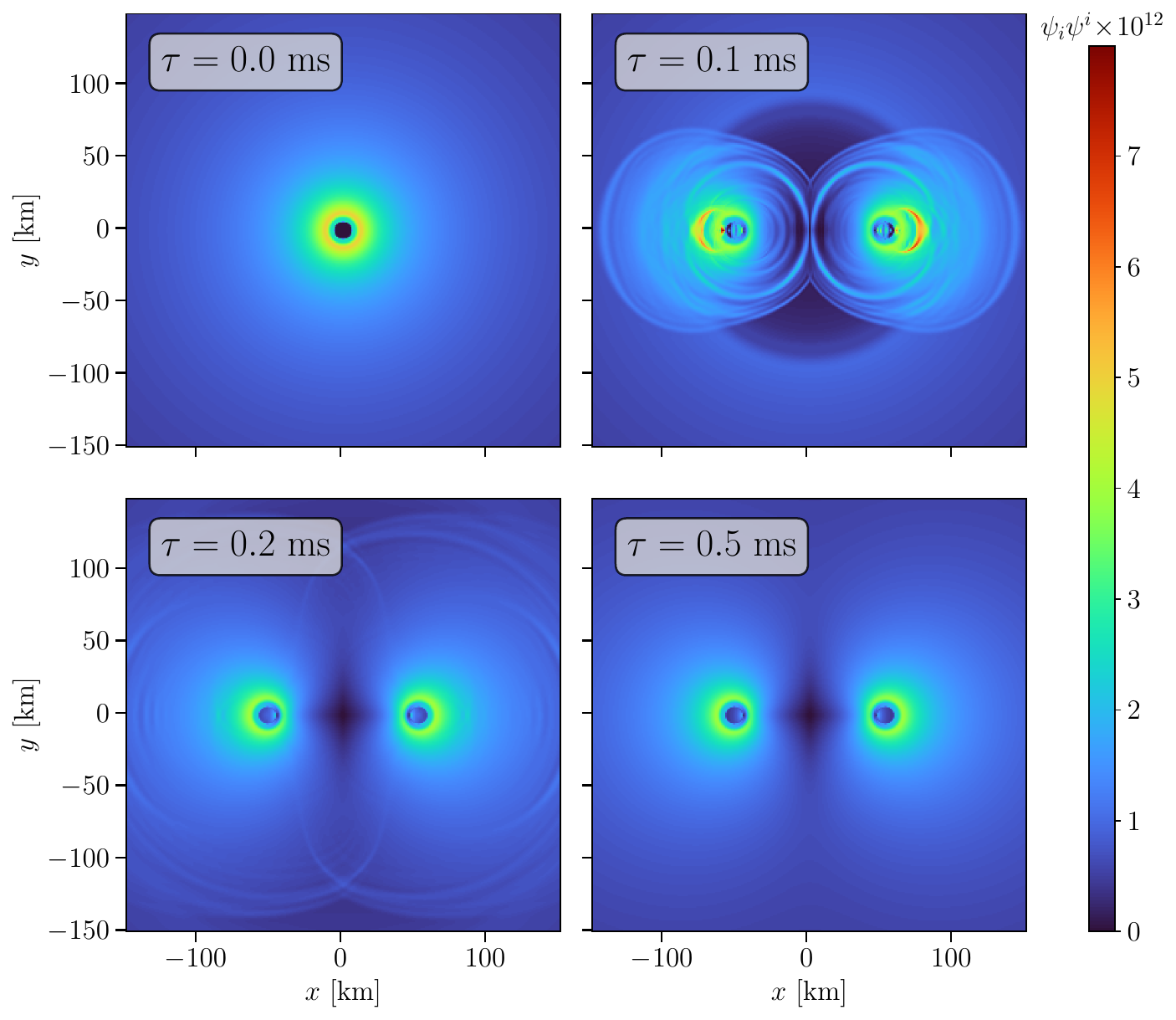}
    \caption{Relaxation of $\psi_{i}\psi^{i}$ under the hyperbolized system of equations, evolving from a single-star initial configuration toward the static binary solution.}
    \label{fig:Hyperbolization}
\end{figure*}

In order to reduce the time it takes until the solution relaxes to the static configuration, we prescribe initial data corresponding to the static solution for a single star with mass equal to the total mass of the binary. This way, the asymptotics of the initial configuration already match those of the binary's static solution. In Fig.~\ref{fig:Hyperbolization} we show the relaxation of $\psi_{i}\psi^{i}$ (i.e.\ derivatives squared of the scalar field) from its initial single-star configuration to the final binary configuration. 

\begin{figure}[h]  
    \centering  
    \includegraphics[width=0.475\textwidth]{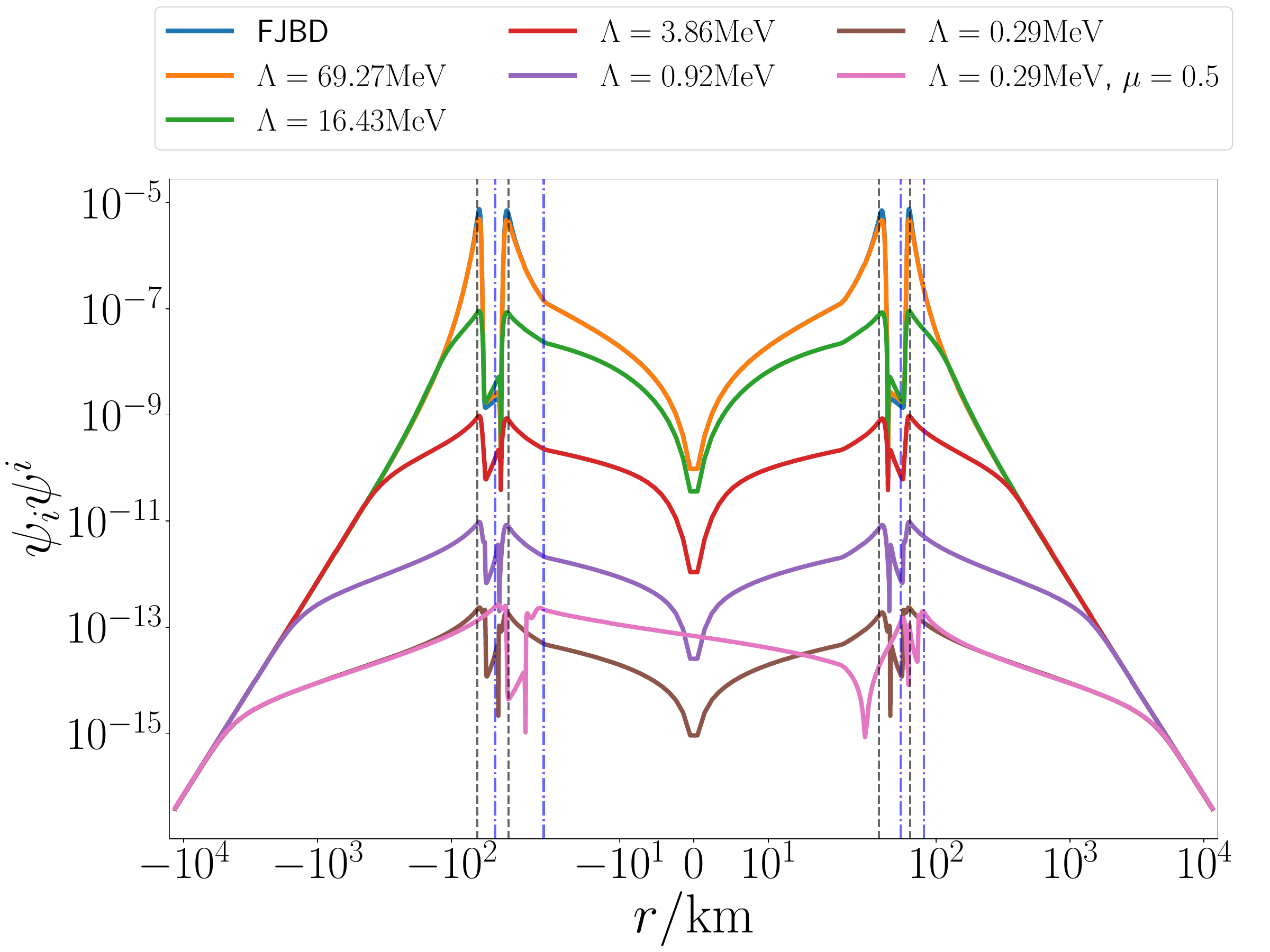}  
    \caption{The scalar field's derivative squared $\psi_{i}\psi^{i}$ across the axis of symmetry, for static solutions with different values of $\Lambda$. In all cases, the total mass is set to $M=2\,M_\odot$ and the radius of each star to $r_s=13.3\,\mathrm{km}$; the stellar interior is delimited by the vertical black dash-dotted line. In the unequal-mass case the second star's interior is indicated by the blue dash-dotted line. The $x$-axis is on a log scale up to $20\,\mathrm{km}$, where it becomes linear.}  
    \label{fig:static_sols}  
\end{figure}

Figure \ref{fig:static_sols} shows the derivative squared of the scalar field $\psi_{i}\psi^{i}$ across the axis of symmetry for different values of $\Lambda$ at fixed mass ratio $\mu\equiv m_{2}/m_{1}=1$, as well as one case with $\mu=0.5$. Here the screening mechanism is clearly observed: as $\Lambda$ decreases, the derivatives of the field are more strongly suppressed and the screening radius (located roughly where the screened solutions match the FJBD solution) increases.

We note that, in order to obtain the solutions presented above, it was necessary to suppress the nonlinear interactions of the scalar field within certain regions inside the star. As mentioned previously, removing the matter source in the stellar interior is sufficient, in the single-star case, to prevent the development of field profiles that would require extremely high resolution to maintain numerical stability. The situation changes in the presence of a companion star. In this case, the scalar field generated by the companion effectively acts as a source within the otherwise empty interior, generating steep profiles that again demand high resolution.
To address this issue, we smoothly suppress the nonlinear interactions of the scalar field within a small region inside the star by introducing a spatial dependence in $1/\Lambda^{8}$, given by:

\begin{equation}\label{eq:LAMBDA}
    \left(\frac{1}{\Lambda^{8}}\right)_{S} = \left(\frac{1}{\Lambda^{8}}\right)\prod_{i=1}^{2}\frac{1}{2} \left( 1 +\tanh{\left(
    \sigma_{S}\left(\frac{\tilde{r}_{i}}{r_{S}}\right)
    \right)}\right), 
\end{equation}
where $r_{S}$ and $\sigma_{S}$ are parameters chosen to adjust the region where nonlinear interactions are turned off. With this prescription we are able to construct the static solutions and keep a stable evolution without altering the solutions of physical observables in the exterior of the stars. In Appendix \ref{app:suppression}, we demonstrate that varying the suppression parameters, such as $r_{S}$ and $\sigma_{S}$, and thus modifying the suppression profile, changes slightly the scalar-field configuration only within the stellar interior. The exterior solution and, consequently, the radiative observables remain unchanged.

 \subsection{Evolution}

Once we have obtained a solution to the static problem, the next step in studying scalar radiation from the binary is to place the system in orbit. To this end, we prescribe a time-dependent orbital angular velocity $\Omega(t)$ for the sources, allowing it to evolve smoothly from zero to its Keplerian value over a timescale $t_{\mathrm{ramp}} = 3.9\,\mathrm{ms}$,

 \begin{equation}
\Omega(t) =
\begin{cases} 
\Omega\displaystyle{\frac{\left(2t_{\rm{ramp}} - t\right)\,t}{t_{\rm{ramp}}^{2}}} & \text{if } 0<t < t_{\rm{ramp}}, \\
\Omega & \text{if }  t \geq t_{\rm{ramp}}.
\end{cases}
\end{equation}

We reduce Eq.~\eqref{eq:scalarEOM} to a first-order system by introducing $\Psi \equiv \partial_{t}\varphi$, together with $\psi_{x} \equiv \partial_{x}\varphi$, $\psi_{y} \equiv \partial_{y}\varphi$ and $\psi_{z} \equiv \partial_{z}\varphi$ (as in the static problem above):
\begin{align}
    &\partial_{t}\left( K^{\prime}(X) \Psi\right) =  \partial_{i}\left(K^{\prime}(X) \psi_{i} \right) -\Sigma~,\\
    &\partial_{t}\psi_{i} = \partial_{i}\Psi~,\\
    &\partial_{t} \varphi = \Psi~.
\end{align}

To study the outgoing scalar radiation, we use the fact that a gravitational-wave detector responds to the Jordan-frame Newman-Penrose invariant $\phi_{22}$, which far from the source can be approximated \cite{Bezares:2021yek,Barausse_2013} as

\begin{equation}
    \phi_{22} = -\alpha \sqrt{16\pi G}\partial_{t}^{2}\varphi + \mathcal{O}\left(1/r^{2}\right)\,.
\end{equation}
This approximation is valid at extraction radii larger than the screening radius $r_{*}$, defined as
\begin{equation}\label{eq:defrs}
    r_* = \bigg( \frac{3}{4} \bigg)^{1/8} \frac{1}{\Lambda} \sqrt{\frac{\alpha M}{4 \pi \mpl}} \; .
\end{equation}

\subsubsection{Equal mass scenario}

We begin by focusing on the equal-mass binary scenario. In this case, symmetry ensures that the scalar radiation is dominated by the quadrupole component. This allows us to study the screening of the quadrupole in isolation from dipole radiation, which would otherwise act as a source for the quadrupole.

For the results presented in this section, each star has a mass $M_{\star} = 1\,M_{\odot}$, and the orbital separation is fixed at $a = 103.39\,\mathrm{km}$. The strong coupling scale $\Lambda$ is varied over the range $0.23\,\mathrm{MeV}$ to $302.85\,\mathrm{MeV}$, and the conformal coupling is set to $\alpha=1/2$.

\begin{figure}[h]  
    \centering  
    \includegraphics[width=0.5\textwidth]{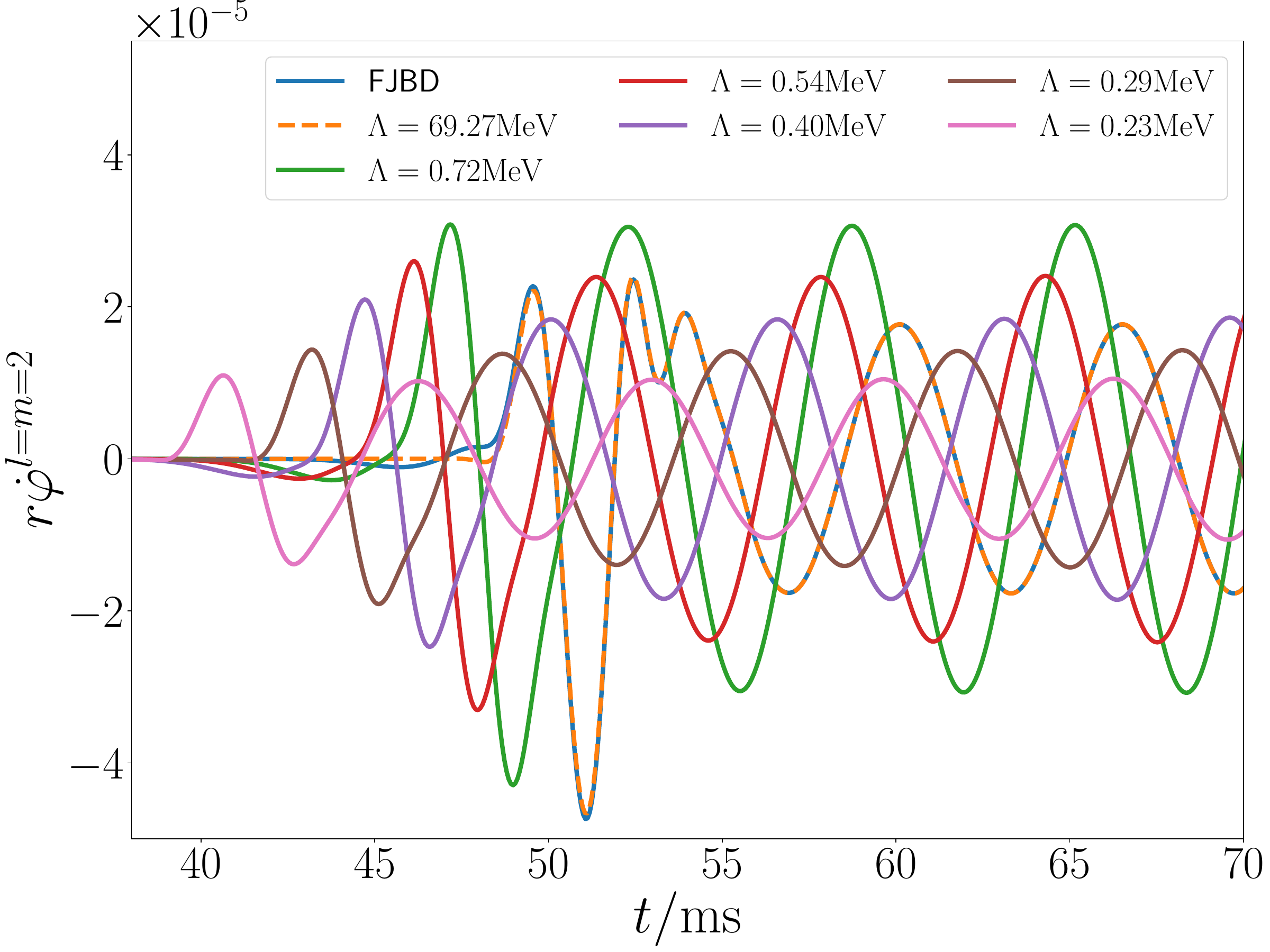}  
    \caption{Waveforms for the quadrupole $\ell=m=2$ mode of the outgoing scalar radiation, for different values of the strong coupling scale $\Lambda$ in the equal mass case. The waveforms are extracted at a radius of $r=7383\,\mathrm{km}$.}  
    \label{fig:waveform_quad}  
\end{figure}

In Fig.~\ref{fig:waveform_quad} we display the quadrupolar $\ell=m=2$ waveforms. We first observe that, after an initial transient associated with the ramp-up of the orbital frequency, the waveforms settle into the stationary behavior consistent with Keplerian orbits. We also observe a phase shift relative to the FJBD waveform; the shift grows as $\Lambda$ decreases. More precisely, the phase shift in the quadrupolar radiation follows $\delta \chi(\varphi^{\ell=m=2}) = 0.78\,\Omega r_{*}$, slightly different from the value $1.12$ entering Eq.~\eqref{eq:homogeneousSolLargeDistance}; this discrepancy suggests an additional phase contribution to the order-one constant $C$ in Eq.~\eqref{eq:amplitudePrediction}.

Strikingly, as $\Lambda$ decreases the amplitude of the radiation initially \emph{increases}, contrary to the naive expectation that it should be progressively suppressed by screening. However, once $\Lambda$ becomes sufficiently small, the amplitude begins to decrease, eventually dropping below the FJBD value. To make this behavior more explicit, we extract the amplitude at several extraction radii for each $\Lambda$. Knowing that for radii larger than the screening radius the amplitude of $\varphi$ behaves like $A_{\varphi}(r) = \mathcal{A}/r + \mathcal{O}(r^{-2})$, we fit for the value of $\mathcal{A}$ for different values of $\Lambda$.

\begin{figure}[h]  
    \centering  
    \includegraphics[width=0.475\textwidth]{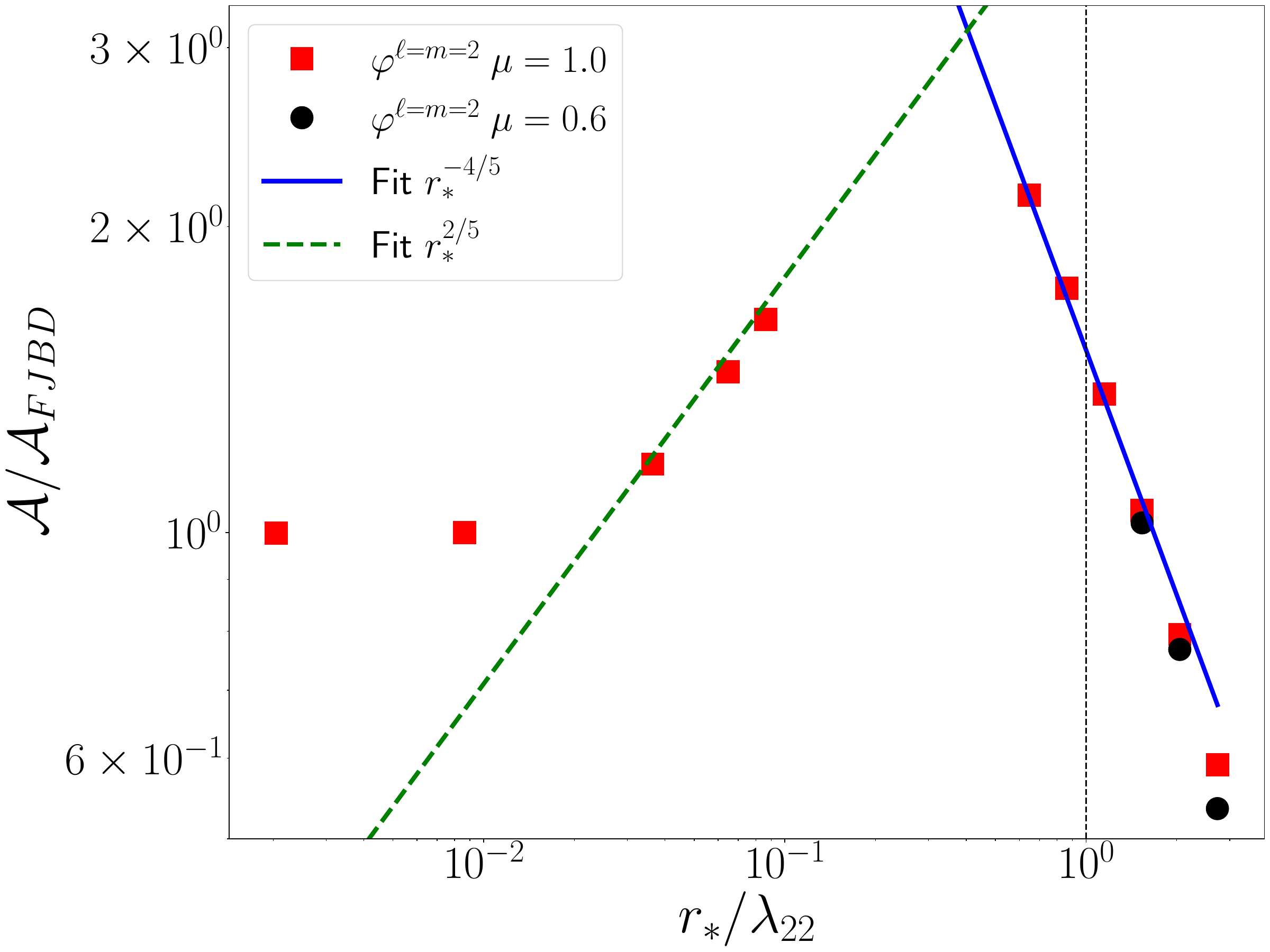}  
    \caption{The quadrupole amplitude $\mathcal{A}$ relative to its FJBD value $\mathcal{A}_{\mathrm{FJBD}}$, as a function of $r_{*}/\lambda_{22}$, where $\lambda_{22}$ is the wavelength of the $\ell=m=2$ radiation. The vertical dashed line marks $r_{*}/\lambda_{22} = 1$.}  
    \label{fig:Amplitude_FJBD}  
\end{figure}

In Fig.~\ref{fig:Amplitude_FJBD} we show the ratio $\mathcal{A}/\mathcal{A}_{\mathrm{FJBD}}$ for different values of $\Lambda$, where $\mathcal{A}_{\mathrm{FJBD}}$ denotes the corresponding value in the FJBD case. We clearly observe that screening becomes significant when the radiation wavelength $\lambda_{22}$ is smaller than the screening radius $r_{*}$, with results consistent with a scaling $\mathcal{A} \propto r_{*}^{-4/5}$. This confirms our analytic argument in section~\ref{sec:expected_scaling}. In contrast, in the regime $\lambda_{22}>r_*$ the behavior is consistent with $\mathcal{A} \propto r_{*}^{2/5}$ and we do not yet have an analytic understanding of this scaling. It was not practical to perform additional simulations in the intermediate regime between these two behaviors to fully resolve the transition. In this region, we found that stable simulations require adjusting the value of $r_{S}$ \eqref{eq:LAMBDA} to properly resolve the steep gradients of the scalar field configuration.

These results are consistent with the full numerical-relativity BNS-merger simulations of Ref.~\cite{Bezares:2021dma} in the $\lambda_{22}>r_*$ regime, and with the perturbative study of scalar modes around a neutron star in $K$-essence of Ref.~\cite{Shibata:2022gec}. We note, however, that the long-wavelength enhancement of the quadrupolar amplitude seen here was \emph{not} reported in the NS-BH simulations of Ref.~\cite{Cayuso:2024ppe}, where the quadrupolar amplitude instead decreased monotonically as $\Lambda$ was lowered, plateauing at a factor $\sim 3$ below FJBD. We tentatively attribute this difference to the presence of a dominant scalar dipole in the NS-BH setup, which is absent in the equal-mass BNS case considered here and may non-linearly couple to the quadrupole.

\subsubsection{Unequal mass scenario}

In this section, we investigate how the scalar radiation depends on the stellar mass ratio. When $\mu<1$, a dipolar contribution emerges, allowing us to assess whether and to what extent it influences the quadrupolar emission.

We focus on binary systems with a fixed value of the parameter $\Lambda = 0.29\,\mathrm{MeV}$ and an orbital separation $a = 103.39\,\mathrm{km}$. We consider mass ratios in the range $\mu \in [0.5, 1.0]$, while keeping the total mass fixed at $M = 2\,M_{\odot}$.

\begin{figure} 
    \centering  
    \includegraphics[width=0.475\textwidth]{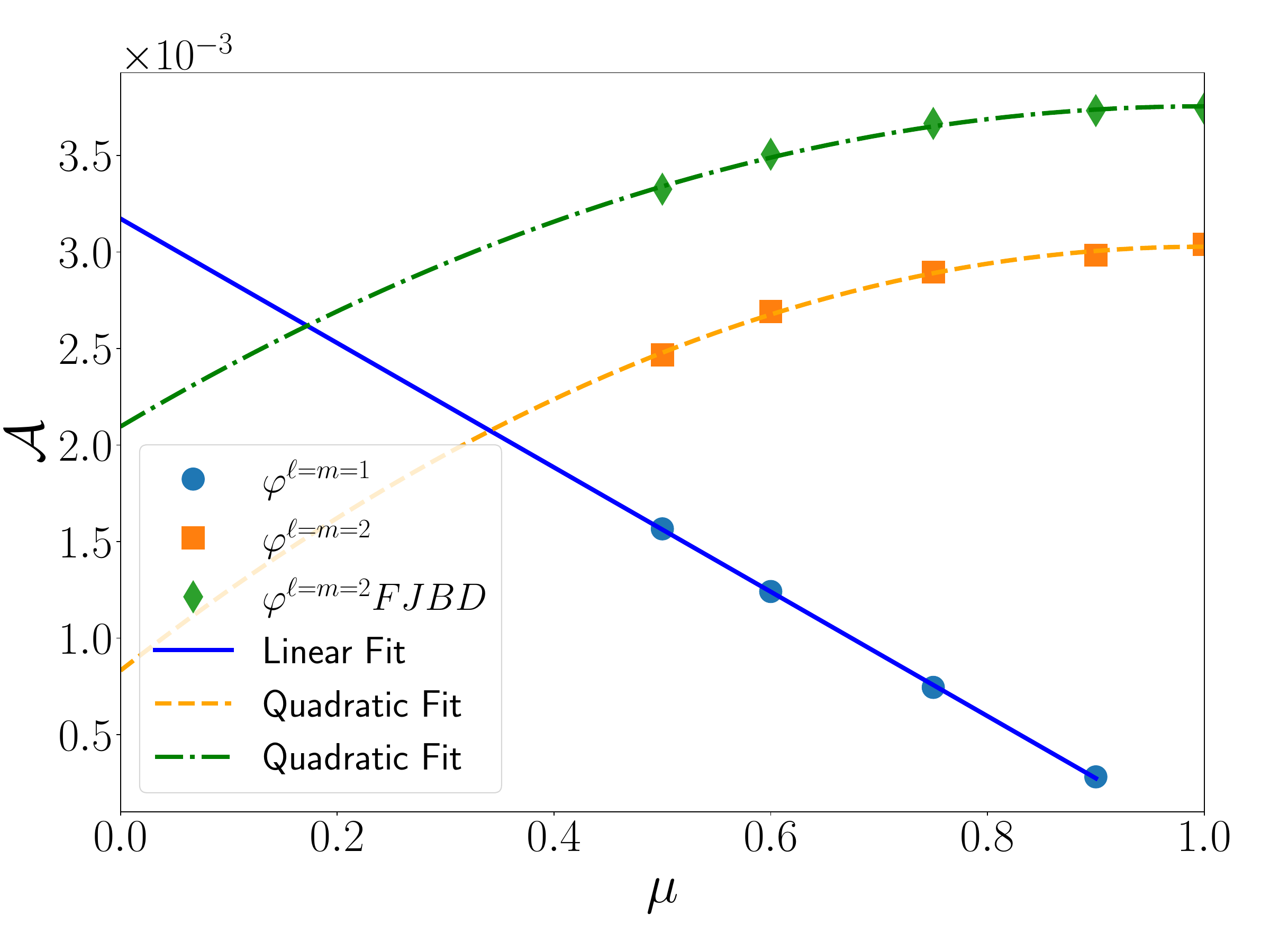}  
    \caption{Amplitude $\mathcal{A}$ of the dipole and quadrupole as a function of the mass ratio $\mu$. All points are obtained for systems with $\Lambda = 0.29\,\mathrm{MeV}$. }  
    \label{fig:Amplitude_mu}  
\end{figure}

In Fig.~\ref{fig:Amplitude_mu} we plot the value of $\mathcal{A}$ for both the dipolar and quadrupolar contributions to the radiation for several values of $\mu$, at fixed $\Lambda = 0.29\,\mathrm{MeV}$, together with the corresponding FJBD case. As $\mu$ decreases, the dipolar contribution $\varphi^{\ell=m=1}$ increases approximately linearly, while the quadrupolar component $\varphi^{\ell=m=2}$ decreases quadratically. For this setup, extrapolating from our results suggests that the dipole becomes dominant over the quadrupole for $\mu \lesssim 0.35$. In this regime, we expect deviations from the clean quadrupolar suppression shown in Fig.~\ref{fig:Amplitude_FJBD}, approaching the dipole-dominated NS--BH regime of Ref.~\cite{Cayuso:2024ppe}. Unfortunately, we were unable to perform simulations in this regime \footnote{As the mass ratio decreases, the influence of the more massive star on its lighter companion becomes increasingly strong. To avoid the formation of steep gradients in the stellar interior, one would need to further tune the parameters controlling the suppression of the nonlinear interactions in Eq.~\eqref{eq:LAMBDA}.}.

The black dots in Fig.~\ref{fig:Amplitude_FJBD} show that for $\mu = 0.6$ the screening behavior in the $\lambda_{22} < r_{*}$ regime remains very close to that observed in the equal-mass case $\mu = 1$. This is expected, since for this value of $\mu$ the dipolar contribution is still approximately half that of the quadrupole and therefore not large enough to significantly affect the quadrupolar emission.

\section{Conclusions} \label{sec:conclusions}

In this paper we have investigated the scalar emission from binary neutron stars in shift-symmetric scalar-tensor theories with kinetic screening ($K$-essence), building on the decoupling-limit approach of Ref.~\cite{Cayuso:2024ppe} and adapting it to a binary configuration in which both compact objects carry a non-vanishing scalar charge (e.g.\ two neutron stars). At the numerical level, a central difficulty in this class of theories is the appearance of Keldysh-type breakdowns in the scalar Cauchy problem. These breakdowns occur, in our setup,
when producing initial data by relaxation from configurations
far from the stationary one (e.g.\ trivial scalar field configurations). We bypass these problems by hyperbolizing the static field equations; the resulting scheme allows for constructing static binary initial data even when the screening radius greatly exceeds the orbital separation---a regime that has so far proved challenging for full numerical-relativity approaches.

To avoid the aforementioned breakdowns, we then smoothly ramp up the orbital angular velocity from these initial data to its Keplerian value, and track the outgoing scalar radiation; in this way we have uncovered a non-monotonic dependence of the quadrupolar scalar amplitude $\mathcal{A}$ on the strong-coupling scale $\Lambda$. For equal-mass configurations, where symmetry forces the dipole to vanish, we find two distinct regimes separated by $\lambda_{22}\sim r_*$: at short wavelengths $\lambda_{22}<r_*$, $\mathcal{A}/\mathcal{A}_\mathrm{FJBD}\propto r_*^{-4/5}$, while at long wavelengths $\lambda_{22}>r_*$ the amplitude is enhanced above the FJBD value, with a power-law fit consistent with $\mathcal{A}\propto r_*^{2/5}$. These two regimes connect prior observations in the literature: the clean short-wavelength suppression of the scalar quadrupole found in perturbative studies of isolated $K$-essence stars~\cite{Shibata:2022gec}, and the long-wavelength enhancement of the scalar quadrupole reported both in those same perturbative studies and in the full numerical-relativity BNS-merger simulations of Ref.~\cite{Bezares:2021dma}. By contrast, the long-wavelength enhancement was \emph{not} observed in the NS-BH simulations of Ref.~\cite{Cayuso:2024ppe}, where the quadrupolar amplitude decreased monotonically with $r_*$; we tentatively attribute this difference to the strong scalar dipole intrinsic to the NS-BH configuration, which is absent at equal masses in a BNS and may non-linearly couple to the quadrupole. Breaking the equal-mass symmetry reintroduces a scalar dipole, whose amplitude grows linearly with the mass asymmetry, while the quadrupolar amplitude decreases quadratically; the quadrupolar screening nevertheless remains essentially indistinguishable from the equal-mass case for mass ratios $\mu\gtrsim 0.6$.

These findings have direct observational implications. The non-monotonic character of kinetic screening---enhancing the signal at low frequencies and suppressing it at high frequencies---shows that the naive picture of a uniform cut-off on scalar emission is incorrect. As a concrete example, the relativistic double pulsar PSR~J0737-3039~\cite{Kramer:2006nb,Kramer:2021jcw} emits $\ell=m=2$ scalar radiation at $\lambda_{22}\sim 10^{9}$~km, while cosmologically-motivated values of $\Lambda$ give $r_*\sim 10^{11}$~km around a solar-mass source. The system therefore lies in the screened region of parameter space, but its quadrupolar amplitude is suppressed only by a factor of a few tens relative to FJBD. The observed $\mathcal{A}\propto r_*^{-4/5}$ scaling at $\lambda_{22}<r_*$ is accounted for by the analytic argument presented in Section~\ref{sec:expected_scaling}.

Several natural extensions are left to future work. The transition region $\lambda_{22}\sim r_*$---where the scaling switches between the two regimes uncovered here---would require further case-by-case tuning of the numerical setup to control the steep gradients that develop in the stellar interior in this regime; pinning down its shape quantitatively is left to a dedicated numerical or semi-analytic study. Moving beyond the decoupling limit to include the backreaction of the scalar on the metric and on the orbital dynamics will also be needed to convert our amplitude scalings into a phase evolution suitable for post-Newtonian waveform modelling---the final step in turning our findings into quantitative bounds on $\Lambda$ from observations.

\acknowledgements

We acknowledge support from the European Union's Horizon ERC Synergy Grant ``Making Sense of the Unexpected in the Gravitational-Wave Sky'' (Grant No.\ GWSky--101167314; to R.C. and E.B.); from the PRIN 2022 grant ``GUVIRP - Gravity tests in the UltraViolet and InfraRed
with Pulsar timing'' (to R.C., E.B. and A.K.). T.A. gratefully acknowledges support from NSF grants OAC-2229652 and AST-2108269. 
A.K. thanks the Fundação para a Ciência e Tecnologia (FCT), Portugal, for the financial support to the Center for Astrophysics and Gravitation (CENTRA/IST/ULisboa) through grant No. UID/PRR/00099/2025 and grant No. UID/00099/2025, as well as to the FCT project ``Gravitational waves as a new probe of fundamental physics and astrophysics'' grant agreement 2023.07357.CEECIND/CP2830/CT0003.
 MB acknowledges partial support from the STFC Consolidated Grant nos. ST/Z000424/1 and UKRI2492. This work used the DiRAC Memory Intensive service Cosma8 at Durham University, managed by the Institute for Computational Cosmology on behalf of the STFC DiRAC HPC Facility (\url{www.dirac.ac.uk}). The DiRAC service at Durham was funded by BEIS, UKRI and STFC capital funding, Durham University and STFC operations grants. DiRAC is part of the UKRI Digital Research Infrastructure.

\appendix

\section{Robustness of the interior suppression prescription}
\label{app:suppression}

As discussed in Sec.~\ref{sec:ID}, we introduce a smooth spatial dependence in the nonlinear coupling,
Eq.~(\ref{eq:LAMBDA}), to suppress the nonlinear interactions of the scalar field within a small
region inside each neutron star. This prescription is introduced solely as a numerical
regularization that allows us to construct the initial data and perform stable evolutions in
situations where large scalar-field gradients develop inside the stellar interior. Since this
modification is not part of the physical model, it is important to verify that our results are
independent of the particular choice of suppression profile.

\begin{figure} 
    \centering  
    \includegraphics[width=0.475\textwidth]{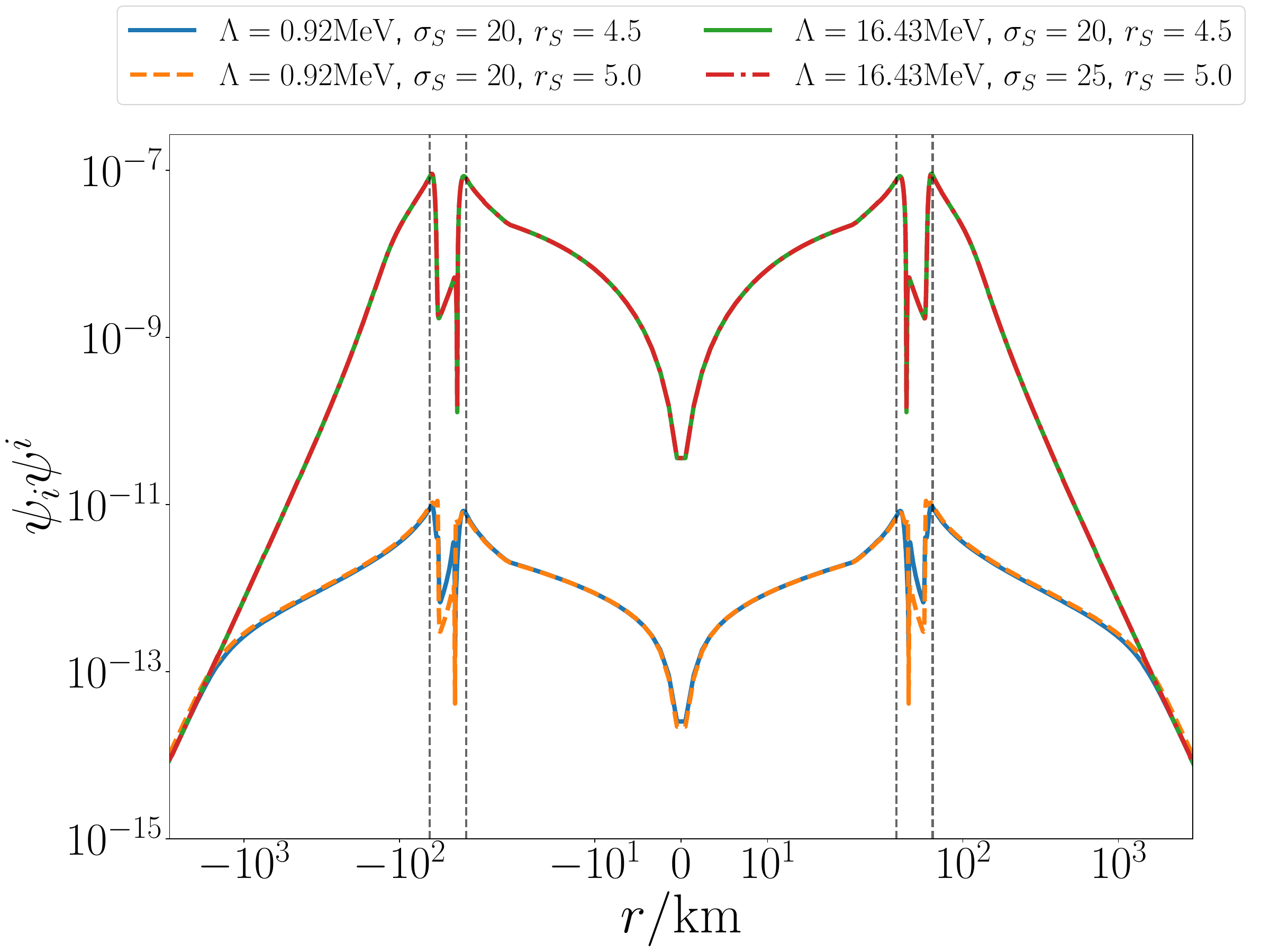}  
    \caption{Initial scalar field's derivative squared $\psi_{i}\psi^{i}$ across the axis of symmetry profile for different choices of the suppression parameters
$(r_S,\sigma_S)$ and two representative values of $\Lambda$. The vertical dashed lines denote the
stellar surfaces. Varying the suppression profile changes the scalar field only inside the stars,
while the exterior solution remains unchanged. The $x$-axis is on a log scale up to $20\,\mathrm{km}$, where it becomes linear. }  
    \label{fig:ID_suppression}  
\end{figure}

To assess the robustness of the prescription, we performed simulations with different values of
the suppression parameters $r_S$ and $\sigma_S$. Figure~\ref{fig:ID_suppression} compares the
initial scalar-field profiles for two representative values of $\Lambda$ using different
suppression profiles. Varying the suppression parameters changes the scalar field only
inside the stars, where the nonlinear interactions are artificially switched off. Outside the
stellar surfaces (indicated by the vertical dashed lines), the scalar-field profiles are
essentially identical, confirming that the suppression does not affect the physical exterior
solution.

\begin{figure} 
    \centering  
    \includegraphics[width=0.475\textwidth]{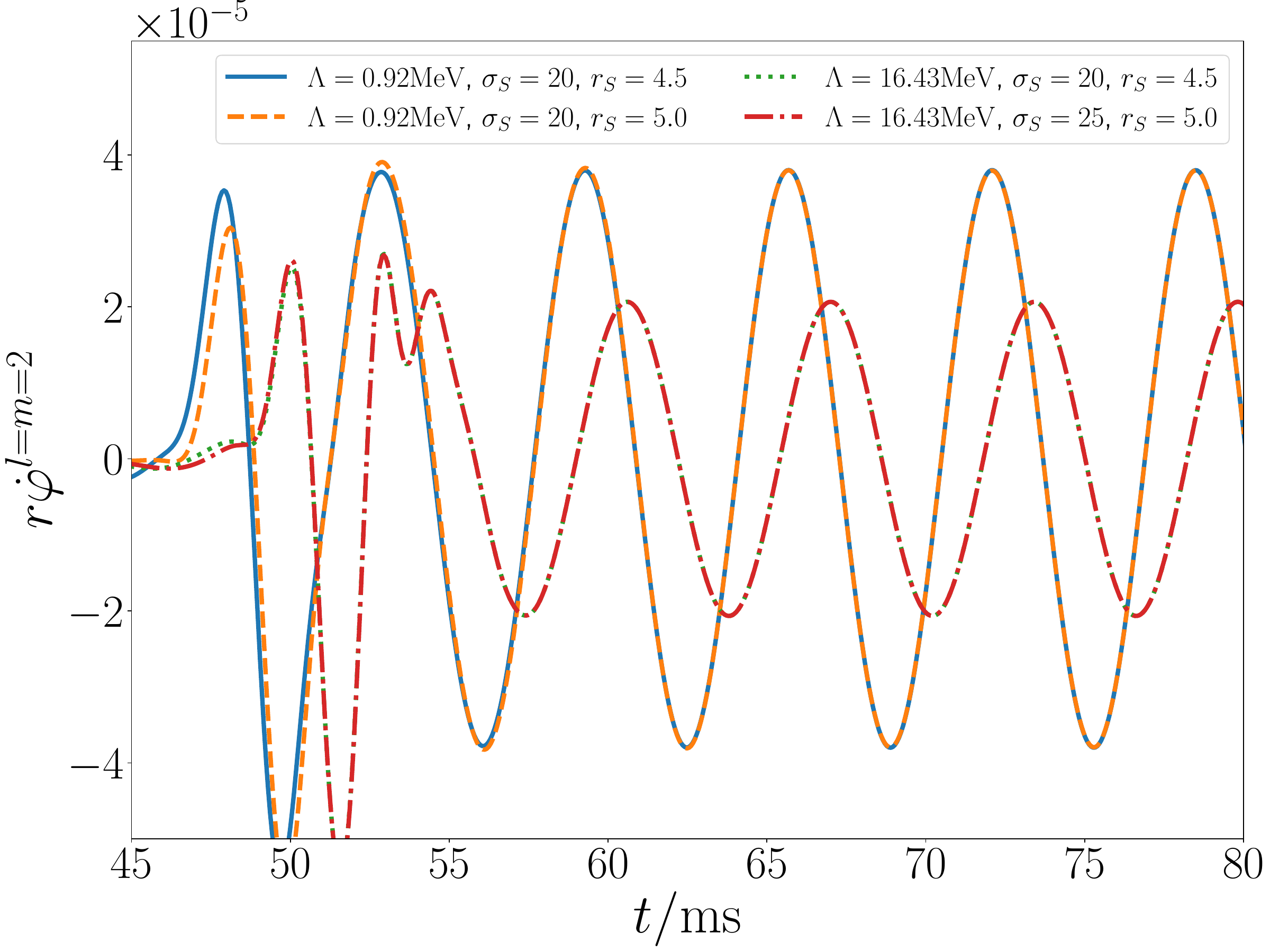}  
    \caption{Dominant $(\ell,m)=(2,2)$ mode of the scalar radiation for different suppression
profiles. The small differences during the initial transient disappear rapidly, and the physical
waveforms overlap throughout the inspiral, demonstrating that the suppression prescription does not
affect the radiative observables.}  
    \label{fig:waveform_suppression}  
\end{figure}

The impact on observable quantities is shown in
Fig.~\ref{fig:waveform_suppression}, where we compare the dominant
$(\ell,m)=(2,2)$ mode of the scalar radiation extracted from simulations using different
suppression profiles. For each value of $\Lambda$, the waveforms overlap almost perfectly after the
initial transient. Small differences are visible only during the early-time junk radiation,
reflecting the modified interior initial data, but these disappear rapidly and do not influence the
subsequent physical evolution. The inspiral waveform, including both its phase and amplitude,
remains unchanged.

These results demonstrate that the suppression prescription acts purely as a numerical
regularization. Although it modifies the scalar-field configuration inside the neutron stars, it
does not alter the exterior solution or any of the radiative observables considered in this work.

\bibliography{bib}

\end{document}